# ELECTRONIC-TRANSPORT PROPERTIES OF Al-Cu-Fe THIN FILMS

R. HABERKERN, C. ROTH, R. KNÖFLER, F. ZAVALICHE*, P. HÄUSSLER
*TU Chemnitz, Inst. für Physik, 09107 Chemnitz, Germany*
\* *permanent address: Inst. of Phys. and Tech. of Materials, Bucharest, 76900 Romania*

Amorphous samples of Al-Cu-Fe were crystallized to the icosahedral phase under the control of electrical conductivity and thermopower at a rather small temperature of $T$=720 K. A comparison between different annealing states of the amorphous and the resulting quasicrystalline phase done on the same sample shows astonishing parallels in conductivity and large differences in thermopower.

## 1. Introduction

Stable quasicrystals show properties which are unexpected for a system of only metallic components. Parts of this can be attributed to the strong interrelation between the special quasicrystalline structure and the electronic system. Generally there are two physical pictures based on apparently different features of the quasicrystalline state. On one hand a Hume-Rothery behaviour is discussed, based on the enhanced electronic stabilisation of the quasicrystalline state by the high multiplicity of the Bragg-scattering vectors building up the relevant nearly spherical Jones-zone[1]. Here, the properties of stable quasicrystals are interpreted by a small density of electronic states (pseudogap) at the Fermi energy $E_F$. Subsequently the compensation of the resulting electron- and hole-like charge carriers arises. This can explain the magnitude and the temperature dependence of transport measurements, especially for band-sensitive effects as thermopower and Hall effect[2]. On the other hand the localization picture is based on the confinement of some conduction electrons inside of clusters. The electronic transport is ruled by the damping of the wave-function on its way to the next identical cluster[3,4].

We report here on electronic transport measurements on thin film samples which were produced in an amorphous state and afterwards crystallized to the icosahedral one.

## 2. Experimental

Amorphous thin film samples of Al-Cu-Fe were prepared by sequential flash evaporation[5] onto sapphire and glass substrates cooled down to $T$=4K. This yields homogeneous films of 20 - 200 nm thickness with a tolerance in composition of typically less than 0.3 at%. Electrical conductivity and thermopower were measured in situ by a 4 probe technique in the temperature range 1.2-350 K. Afterwards the samples were annealed in a different apparatus under the control of conductivity and thermopower with constant heating rates (1-5 K/min) in the temperature range 300-900 K (vacuum $p < 1*10^{-7}$ mbar at 900 K). The structure factor of the amor-phous phase was measured by electron diffraction on free standing films. The icosahedral phase was determined by electron diffraction in a commercial TEM.

## 3. Results and Discussion

A typical plot of an in situ measurement of the electrical resistivity beginning with the as-evaporated film at $T=4K$ ($\rho_0(5K)=300$ μΩcm) is shown in fig 1a. The amorphous film (thickness 65 nm) was thermally cycled between stepwise increasing temperatures of 90, 200, 350, 500, 700 K and 1.3 K, respectively. This allows to separate the irreversible from the reversible behaviour. The irreversible increase of the resistivity is attributed to a relaxation process of the amorphous structure where, with increasing temperature, the atoms have the possibility to move into the energetically favourable sites of the Friedel minima. This results in a stronger interrelation between ionic structure and the electronic system. The resistivity increases by a stronger elastic umklapp scattering of the conduction electrons (described by the Fermi vector $k_F$) at a strong peak in the structure factor at $G=2k_F$. The most remarkable feature in fig. 1a is the sharp increase of resistivity at $T=760$ K followed by a temperature dependence of the resistivity characteristic for high quality quasicrystalline samples in the Al-Cu-Fe system. The transition from the amorphous to the icosahedral phase was also proved by electron diffraction. The transition occurs at a remarkably low temperature, about 300 K lower than the annealing

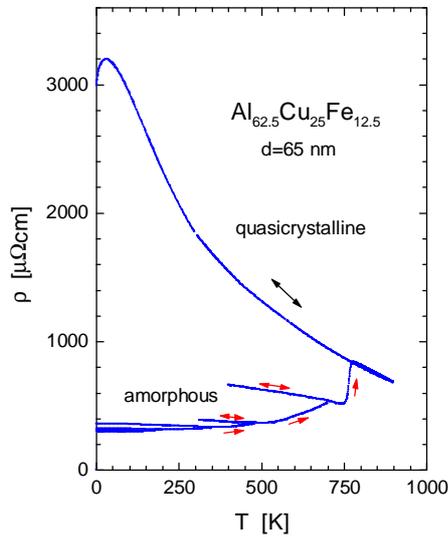 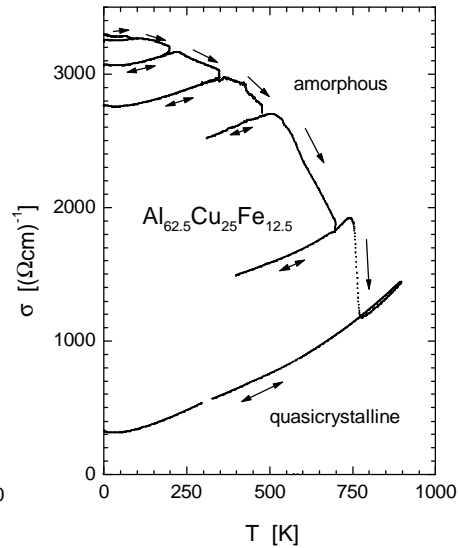

fig. 1a: resistivity       fig. 1b: conductivity

as a function of temperature for an amorphous prepared sample, stepwise heated up and crystallised to the icosahedral phase at T= 760 K

temperatures necessary to remove crystalline inclusions in quasicrystalline bulk samples. It is also much smaller than the temperatures necessary to form the icosahedral phase in films made by consecutive sputtering of one-component layers[6]. We explain this by a probably similar short range order between the amorphous and the icosahedral phase, so only short-range diffusion is necessary to form the quasicrystal. It is an advantage of the method used here, that the metastable amorphous state (produced at low temperatures) has a liquidlike order but the dynamics of any structural change itself is governed by the temperature given by the experimental condition. So it is possible to see the transition from the disordered (amorphous) phase to the quasicrystalline state already at $T \approx 700$ K. The transition for the disordered liquid to the quasicrystalline phase occurs at $T \approx 1100$ K and on a time scale decreased by many orders of magnitude because of the high temperature. This different time scale also influences the dynamics in the constitution of phases like the cubic phases which generally occur in Al-Cu-Fe samples prepared by melt-spinning. A heating rate of 5 K/min applied to the amorphous samples avoids the formation of a larger portion of a cubic phase.

While the discussion of the resistivity is usual for amorphous metals, quasicrystals are frequently discussed in a picture of electrical conductivity because of their *anti-metallic* behaviour. Fig. 1b shows the same data as fig. 1a but replotted for the conductivity. The lowest curve shows the characteristic behaviour of a good

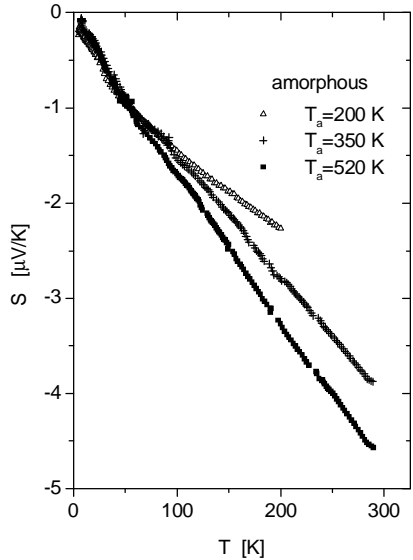 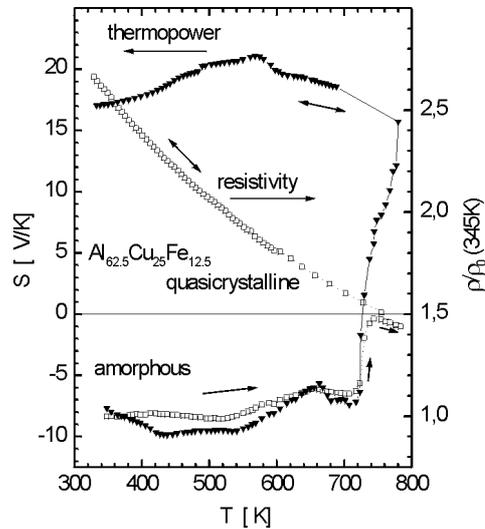

fig. 2: reversible part of the thermopower for amorphous samples annealed at $T_a$=200, 350 and 520 K. (curve for $T_a$=200 K measured by C. Lauinger)

fig. 3: thermopower and normalised 2 probe resistivity (measured with the same leads as the thermopower) for an amorphous evaporated sample showing a transition to the icosahedral phase at 730 K

quasicrystal whereas the upper reversible parts of the curve which belong to the amorphous state in different annealing steps are nearly parallel to the quasicrystalline behaviour. Such a behaviour was reported many times for quasicrystals[7,2] depending on composition and structural quality. This was attributed to a conductivity which is composed of a low temperature value characteristic to composition and structural quality plus a temperature dependant increase in conductivity similar for all quasicrystals. Astonishingly the differently annealed amorphous states show exactly the same behaviour as the quasicrystal developed from it. Furthermore the crystallization to the icosahedral phase looks less like a qualitatively new feature than like the succession of the irreversible relaxation processes in the amorphous state at lower temperatures.

The behaviour of the thermopower (fig. 3) is quite different. At the amorphous to quasicrystalline transition the thermopower changes from a small negative to a large positive value while the different annealing steps of the still amorphous phase plotted in fig. 2 show more negative values for the samples annealed at higher temperatures. This could be interpreted as a hint for a different band-structure of amorphous and quasicrystalline phase in the Hume-Rothery picture for which the thermopower has a higher sensitivity than the conductivity. A detailed discussion can be found elsewhere[8].

### 3. Conclusion

We have shown that it is possible to produce very thin quasicrystalline films in a quality unexpected for that thickness via the amorphous phase. The in situ measurement of conductivity and thermopower allows us to optimize the annealing procedure and to compare the amorphous to the quasicrystalline phase on the same sample.

### Acknowledgement

We would like to thank DFG for its support under contract number Ha 2359/1 and Ha 2359/2.